# Filtration and breathability of nonwoven fabrics used in washable masks

Thomas W. Bement[1], Ania Mitros[3], Rebecca Lau[2], Timothy A. Sipkens[1,2], Jocelyn Songer[2], Heidi Alexander[1], Devon Ostrom[4], Hamed Nikookar[1], Steven. N. Rogak[1,*]

[1] Department of Mechanical Engineering, University of British Columbia, Vancouver BC Canada
[2] Metrology Research Center, National Research Council of Canada, Ottawa ON Canada
[3] MakerMask, Orange, Massachusetts, *https://makermask.org*
[4] Artist/Researcher, Toronto, ON, *https://ostrom.ca*

[*] Corresponding author. Tel: 1-604-822-4149; Fax: 1-604-822-2403
E-*mail address*: rogak@mech.ubc.ca

## Abstract

This study explores nonwoven and woven fabrics to improve upon the performance of the widespread all-cotton mask, and examines the effect of layering, machine washing and drying on their filtration and breathability for submicron and supermicron particles. Individual materials were evaluated for their quality factor, $Q$, which combines filtration efficiency and breathability. Filtration was tested against particles 0.5 μm to 5 μm aerodynamic diameter. Nonwoven polyester and nonwoven polypropylene (craft fabrics, medical masks, and medical wraps) showed higher quality factors than woven materials (flannel cotton, Kona cotton, sateen cotton). Materials with meltblown nonwoven polypropylene filtered best, especially against submicron particles. Subsequently, we combined high performing fabrics into multi-layer sets, evaluating the sets' quality factors before and after our washing protocol, which included machine washing, machine drying, and isopropanol soak. Sets incorporating meltblown nonwoven polypropylene designed for filtration (Filti and surgical mask) degraded significantly post-wash in the submicron range where they excelled prior to washing ($Q$ = 57 and 79 at 1 μm, respectively, degraded to $Q$ = 10 and 15 post-wash). Washing caused lesser quality degradation in sets incorporating spunbond non-woven polypropylene or medical wraps ($Q$ = 12 to 24 pre-wash, $Q$ = 8 to 10 post-wash). Post-wash quality factors are similar for all multi-layer sets in this study, and higher than Kona quilting cotton ($Q =$ 6). Washed multi-layer sets filtered 12% to 42% of 0.5 µm, 27% to 76% of 1 µm, 58% to 96% of 2.8 µm, and 72% to 100% of 4.2 µm. The measured filtration and pressure drop of both the homogeneous and heterogeneous multi-layer fabric combinations agreed with the estimations from the layering model.

# 1 INTRODUCTION

## 1.1 Motivation

Airborne respiratory pathogens, including the SARS-CoV-2 virus responsible for the COVID-19 pandemic and had killed millions of people worldwide annually (Lewis, 2020; Morawska and Cao, 2020; Centers for Disease Control and Prevention, 2021b). Airborne transmission occurs via virus-laden particles generated from breathing, talking, coughing, and sneezing. Research during the 1918 pandemic (Kellogg and MacMillan, 1920) established that masking, including non-medical masks, reduces viral transmission rates (Chu et al., 2020; Eikenberry et al., 2020; Leffler et al., 2020; Lyu and Wehby, 2020; Stutt et al., 2020; Abboah-Offei et al., 2021; Gandhi and Marr, 2021). Current case studies suggest over 70–80% reduction in SARS-CoV-2 transmission rates when masks are used effectively (Doung-Ngern et al., 2020; Malone, 2020; Wang et al., 2020). Public health organizations widely implemented mask mandates to reduce viral transmission rates for the COVID-19 pandemic (World Health Organization, 2020; Centers for Disease Control and Prevention, 2021a). As a result of these mandates and broader public interest in masks, the mask market has expanded with buyers using selection criteria beyond filtration efficiency, including comfort, fashion, environmental impact, cost, access, and supply chain ease.

Washable masks remain widely used, such that it is critical to improve their effectiveness. In this work, we evaluate candidate materials for reusable masks, targeting materials that outperform cotton alternatives in terms of reducing airborne disease transmission. While woven materials, including cotton, are commonly used in washable masks, nonwoven materials, especially meltblowns, have significantly higher quality factors (Maher et al., 2020; Pei et al., 2020; Wilson et al., 2020; Bagheri et al., 2021; Drewnick et al., 2021; Rogak et al., 2021). Li et al. (2020) found comparable performance for cellulose, nonwoven materials, and surgical masks. Nonwoven materials are widely available and commonly used for sewing, crafting, medical sterilization wraps, scrubs, and medical masks. Many are hydrophobic and biocompatible. In this work, a focus is placed on the effect of layering, machine washing, and drying on these materials, as well as composite masks formed from a combination of these materials.

Studies on the effect of mask washing show varied results. Sankhyan et al. (2021) noted the deconstruction of cotton fibers, using electron microscopy, and an increase in inhalation resistance, but no change in filtration efficiency. Hao et al. (2021) found negligible effects of washing on the filtration efficiency of a number of woven materials and a synthetic microfiber cloth. By contrast, samples of N95 mask materials exhibited a reduction in the particle filtration efficiency (PFE) from 98% to 50% for 0.3 µm particles, a likely consequence of a loss of their electret properties. Reutman et al. (2021) considered the effect of washing on a 3-layer mask prototype containing a layer of meltblown polypropylene and found that the filtration efficiency reduced from 85% to 70% for low face velocities and submicron particles. Everts et al. (2021) showed that high-quality medical masks reprocessed 10 times by water immersion methods maintained higher filtration efficiency than new, non-medical, 3-ply disposable masks as well as cotton and cotton-polyester mix fabrics, even when triple layered.

In this work, we present results for single layers of candidate materials, before continuing on to consider the effect of washing and layering, to form composite masks. We use the interim guidance of the World Health Organization (WHO) (2020) for non-medical fabric masks (suggesting a minimum of three layers, including a hydrophilic material for the skin-touching layer, a filter layer, and a hydrophobic outer layer) to guide the formation of composite masks.

**1.2 Background**

Filtration of particles is influenced by four mechanisms: diffusion, impaction, interception, and electrostatic forces (Hinds, 1999). For small particles (<0.1 μm), diffusion plays a prominent role. Larger particles (>0.5 μm) are filtered mostly by impaction and interception. Electrostatic forces are especially influential in the 0.1–2 μm range and key to electret materials. Degradation of electret media by different cleaning methods reduces the quasi-static charge and, thereby, filtration (Xiao et al., 2014; Ou et al., 2020).

Particle size is an important parameter in filtration testing. We tested filtration of particles with aerodynamic diameters between 0.5 μm to 5 μm for several reasons. Firstly, most common fabrics can easily remove particles above >5 μm (Leith et al., 2021) so evaluation at larger sizes does not help differentiate between good candidate materials. Secondly, tests at this size range can be done with a single particle spectrometer, simplifying the experiments. Finally, and most importantly, particles in this size range seem to carry highest risk of infection, as discussed further below.

Bioaerosols generated by breathing, coughing, sneezing, and talking show particle number distribution that peaks around 0.8 μm diameter particles, with numbers of larger particles decreasing up to 1000 μm diameter (Morawska et al., 2009; Johnson et al., 2011; Asadi et al., 2019; Bake et al., 2019). However, viral concentration (copies per unit particle mass) varies with particle size and tends to be highest in particles below 5 μm aerodynamic diameter. A study of COVID-19 patients (Coleman et al., 2021) found that particles ≤5 μm contributed 85% of the total viral RNA load detected from 13 patients. A similar result was observed in monkeys, with 0.65-4.7 μm particles accounting for 77-79% of total virus shed by infected cynomolgus macaques (Zhang et al., 2021). In influenza patients, Milton et al. (2013) found 8.8 times more viral copies in ≤5μm particles than in >5μm particles. In cough droplets of influenza patients, Lindsey et al. (2010) found that 42% of the influenza RNA was contained in particles <1 μm aerodynamic diameter, 23% in particles 1-4 μm, and 35% in particles >4 μm. Looking at transmission rather than viral counts, Zhou et al. (2018) found that droplets <1 μm did not cause ferret-to-ferret influenza transmission, whereas droplets 1.5-15.3 μm did result in infection. Zhou et al. also showed that high particle counts may not imply high infectivity: while 76.8% of total airborne particles released from the ferrets had aerodynamic diameters of 0.52-1.54 μm, ferrets exposed to <1.5 μm particles did not get sick and viral RNA was detectable only in particles >4 μm.

At the same time, where and whether particles settle also depends on particle size. Inhaled particles in the 0.1–10 μm range deposit in the lungs with higher deposition rate for >1 μm particles than <1 μm particles (Park and Wexler, 2008). According to Carvalho et al. (2011), 1-5 μm particles are deposited deep in the lungs, whilst those >10 μm are generally deposited in the oropharyngeal region, and most particles 0.1-1 μm are exhaled. Sosnowski (2021) estimated 75% of particles between 0.2 and 0.8 μm are exhaled and therefore are not deposited in the lungs.

The widely used NIOSH 42 CFR Part 84 standard (hereafter referred to as the NIOSH N95) focuses on filtration efficiency at 0.3 μm, the most penetrating particle size. This standard aims to protect against workplace hazards, like dust, rather than specifically targeting transmission of airborne disease. Medical masks are tested for bacterial filtration efficiency (BFE) with a mean particle size of 3 μm, within our test range.

Particle filtration efficiency (PFE) is defined as

$$\eta = 1 - P = 1 - \frac{N_F}{N_S}, \qquad (1)$$

where η is the PFE, $P$ is the penetration, and $N_f$ and $N_s$ are the filtered and source concentrations, respectively. In terms of source control $N_s$ corresponds to particles generated by an infected individual, while for personal protection $N_s$ corresponds to particles from some external source. The pressure drop ($\Delta p$) across the mask affects both breathing effort and leakage around the sides. Breathing effort directly impacts comfort and thus may impact how consistently a person wears a mask. Quality factor ($Q$), which is a key metric that combines PFE and pressure drop (Hinds, 1999), defined as:

$$Q = \frac{-ln(P)}{\Delta p} = \frac{-ln(1-\eta)}{\Delta p}. \qquad (2)$$

We note that the literature varies on whether the natural or base 10 logarithm is used (Zangmeister et al., 2020).

Within this work, we consider the effect of layering various materials. When filtration is not strongly influenced by particle charge, the filter layers are expected to act independently. Then, for each size class, the net penetration for a layered mask is given by the product of the penetration for the individual layers (a consequence of each layer seeing only the particles not filtered by the previous layers),

$$P_{tot} = \prod_j P_j , \qquad (3)$$

and, thus,

$$\eta_{tot} = 1 - \prod_j (1 - \eta_j) , \qquad (4)$$

where $P_{tot}$ and $\eta_{tot}$ are the penetration and PFE for the composite mask and $P_j$ and $\eta_j$ are the penetration and PFE for the $j$th layer. (We note that, integrated filtration efficiencies, such as the mass-based filtration efficiency targeted by the TSI 8130A, are less likely to follow this trend, due to the nature of the integration step.) By contrast, the pressure drop is given by the sum of each layer, as an extension of Darcy's law. As a result, any single material is expected to have the same quality factor, regardless of the number of layers.

## 2 METHODS

**2.1 Apparatus**

A TSI 3076 atomizer generated the challenge aerosol from a 20g/L NaCl solution. The atomizer was fed with air by a mass flow controller (ALICAT MCS-10SLPM-TFT) supplied with pressurized room air at 2-3 SLPM. Total particle concentration was below 3000 #/cm$^3$, according to the maximum concentration range of the TSI 3330, and was not adjusted during the tests. A Senserion SPS30 was used to monitor the upstream to ensure that aerosol concentrations were consistent. The particles were diluted with room air in an extraction duct (ambient particles were less than 1% of the total).

The NaCl particles passed through an x-ray charge neutralizer (TSI model 3088) resulting in a quasi-equilibrium bipolar charge distribution (Johnson et al., 2011). The effect of this distribution was checked using a differential mobility analyzer (DMA) column as an electrostatic precipitator to remove all of the charged particles. Comparing the neutralized to uncharged case, we confirmed that the variation of net charge in the neutralized distribution was negligible (see Supplemental Information A; we note that, unlike Corbin et al. (2021), we are not considering electret materials and are considering larger particles). The remaining tests were run without the DMA column. Following neutralization, the aerosol was diverted to either a bypass line or through flat filtration media that was clamped in a holder with a 21 mm diameter flow passage.

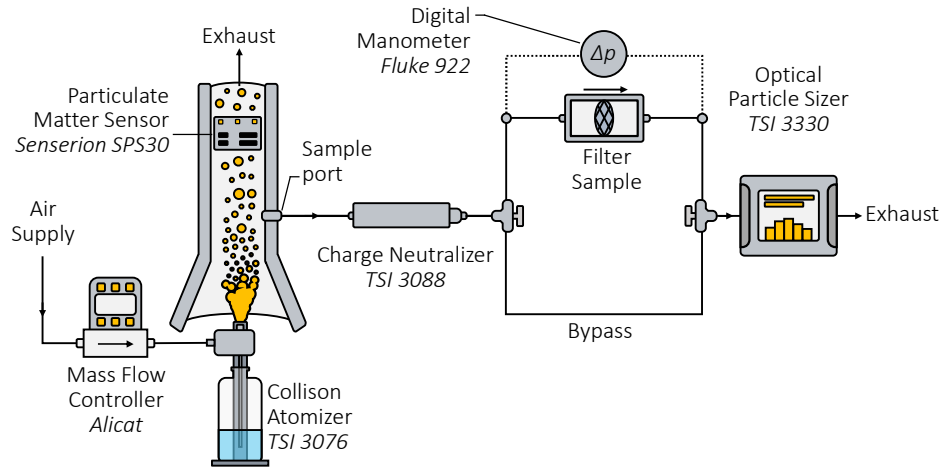

**Fig. 1.** Filter test apparatus. The duct between the atomizer and the sample port was approximately 2 meters. The sample was drawn from near the midpoint of this duct.

Following the sample/bypass section, an Optical Particle Sizer (OPS, TSI Model 3330) measured the total particle counts in 16 size bins ranging from 0.3 to 10 μm (optical equivalent diameter). Filtration efficiency was determined by comparing the counts for the filter versus bypass flow paths. The OPS flow rate was 1 L min$^{-1}$, resulting in a theoretical face velocity (i.e, normal to the fabric) of 4.9 cm s$^{-1}$ through the sample material. Considering typical mask flow areas, this corresponds to an inhalation flow of ~50 L min$^{-1}$, midway between a resting rate and that used for N95 testing (Caretti et al., 2004). Higher face velocities would yield larger PFEs for particles above several microns due to increased impaction, but lower PFE for the smallest particles captured by electrostatic interactions (Corbin et al., 2021).

The OPS sizes particles by the magnitude of the scattered laser light from single particles, assuming a refractive index of 1.4 for NaCl. Using the mean particle size based on the bin limits, we can use this refractive index to convert the bin's mean geometric diameter to the aerodynamic diameter, which is often used when reporting PFE. A table showing the conversion for the bins is included in Supplemental Information B. In what follows, we present PFE as a function of aerodynamic particle size, or for compactness, at an aerodynamic size of 1 micron.

**2.2 Base materials and composite mask**

Nonwoven materials are produced by mechanical, chemical, thermal, or solvent treatments to hold fibre webs together in a disordered matrix. Nonwoven fabrics are either dry formed or wet laid. Dry formed materials are subdivided further into air laid, dry laid, spunbond (Figure 2a), meltblown (Figure 2b), and electrospun (Figure 2c). The nonwoven materials tested in this paper are spunbond (SB), spunbond-meltblown-spunbond (SMS), and electrospun. The SMS materials utilize two SB layers as a substrate and support for the weaker but high filtration meltblown middle layer. Some SMS materials contain additives and coatings which make them ill-suited for masks.

In contrast, woven materials (Figure 2d and 2e) typically have larger gaps between fibers and lower quality factors than nonwovens. However, woven cottons or cotton blends are hydrophilic and are effective as the skin-touching layer of a composite mask.

The individual materials tested in this study are listed, along with their shorthand names and key properties, in Table 1. Each material was tested 3 times with fresh samples in each test. For a subset of the materials, multiple layers were tested together. Based on the measured quality factors, seven composite material sets (A-G, Table 2) were selected as potential improvements over the cotton-only mask and tested.

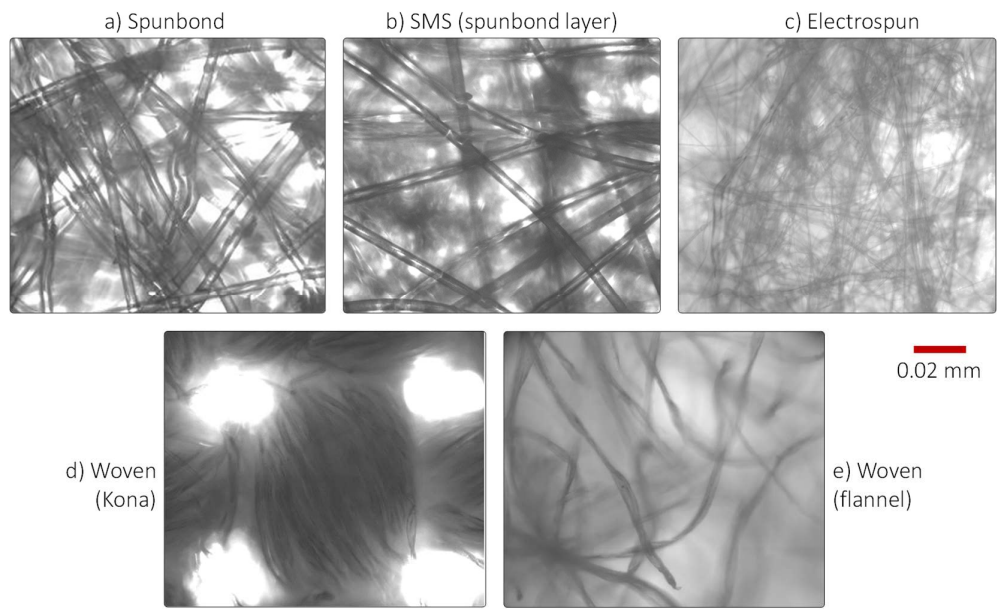

**Fig. 2.** Optical microscopy images of (a) spunbond (SmFb), (b) SMS (H100), (c) electrospun (MaT2), and (d-e) woven (Kona, Flan) materials. The SMS sample shows only the outer spunbond layer as the meltblown layer is hidden below it.

**Table 1.** Set of materials tested in this work, along with their key properties. Quality is taken from Figure 3, based on PFE at 1 μm aerodynamic diameter.

| Name | Short Name | Fibre Material | Manufacturing Method | Weight [g/m$^2$] | Thickness [mm] | Quality [kPa$^{-1}$] |
|---|---|---|---|---|---|---|
| SmartFab Thick | SmFb | Polypropylene | Spunbond | 65 | 0.50 | 10.7 |
| Oly Fun | OlFn | Polypropylene | Spunbond | 69 | 0.50 | 10.5 |
| Pellon 930 | P930 | Polyester | Proprietary** | 41 | 0.18 | 8.6 |
| Advancheck SMS Barrier Autoclave Wrap | Advn | Polypropylene | Spunbond-meltblown-spunbond (SMS) | 45 | 0.26 | 13.6 |
| Halyard material, H400 | H400 | Polypropylene | SMS | 60 | 0.50 | 36 |
| Halyard material, H100 | H100 | Polypropylene | SMS | 34 | 0.23 | 35 |
| Kona | Kona | Cotton | Woven | 153 | 0.50 | 5 |
| Sateen | Satn | Cotton | Woven | 81 | 0.29 | 4 |
| Flannel | Flan | Cotton | Woven and brushed | 176 | 0.58 | 8.3 |
| Filti mask material† | Flti | Polyester-Nanofiber-Polypropylene | Spunlace - electrospun-Spunbond | 76 | 0.47 | 29 |
| N95 3M 1860* | 1860 | Polyester-Polypropylene-Polypropylene | Meltblown (middle layer)*** | 375 | 2.22 | 27 |
| CAN95 IPA* | CN95 | Polypropylene, cotton | Spunbond and meltblown | 205 | 1.34 | 53 |
| Type 2 Red Cross Surgical Mask* | MaT2 | Polypropylene | SMS | 75 | 0.47 | 119 |
| Halyard Mask | HMa | Cellulose, polypropylene | Meltblown (middle layer)*** | 49 | 0.23 | 80 |

\* These materials were treated with isopropanol (IPA) to remove transient static charge.

† Filti was tested both washed and unwashed as an investigation into the specific material. This was a unique test case.

\*\* We suspect this is dry-laid non-woven.

\*\*\* Manufacturing information only available for the meltblown nonwoven layer.

**2.3 Cleaning procedure**

The composite samples (A-G) were also tested before and after ten cleaning cycles to simulate reuse of the masks. Cleaning cycles included machine washing and drying. Samples were washed with a Huebsch commercial front-load washer (HFNLYRSP111CW01) set for "normal loads", "warm water", and "light soil" using a Purex detergent (labeled as "Purex Dirt Lift Action Coldwater Laundry Liquid"). Samples were placed in a laundry bag prior to washing and stayed in the bag through drying. Samples were dried with a Huebsch commercial electric dryer (HDEY07WF1502), set for 60 minutes on "low temperature, rapid". No dryer products were used. To remove transient static charge introduced by the dryer after the 10 machine wash/dry cycles, the samples were soaked in isopropanol (IPA) for at least 6 hours (Xiao et al., 2014), then hung dry for at least 24 hours.

## 3  RESULTS AND DISCUSSION

**3.1 Individual samples**

The materials fall into families with similar quality factors. Figure 3 shows filtration efficiency for 1 μm particles as a function of pressure drop, overlaid with lines of constant quality. Measurements for other particle sizes are included in the online Supplemental Information section

C. Quality factors for the various materials are also given in Table 1. The highest performing materials were the N95 respirators and medical masks, even after IPA treatment, consistent with Rogak et al. (2021), with quality factors from 27 to 120 $kPa^{-1}$. This was followed by the SMS medical wraps (H400, H100, and Advancheck SMS), with quality factors from 13 to 36 $kPa^{-1}$, and the craft-grade spunbond materials (Pellon 930, SmartFab, OlyFun), with quality factor from 8 to 11 $kPa^{-1}$. The Filti mask material performed worse than expected with only 15% filtration at 0.5 μm, while the manufacturer claims 95% filtration at 0.3 μm and 5.33 cm/sec face velocity.

The cotton materials showed a broad range of quality factors depending on the weave. Flannel had the highest quality factor among the hydrophilic woven cotton materials, 8.3, consistent with Zangmeister (2020), who noted higher filtrations in heavily napped cotton fabrics, including flannel. The flannel in this study has a relatively disordered and fluffy structure, visually similar to that of the non-woven materials in this study (see Fig. 2). The low nap woven cottons (Kona and sateen) exhibited the lowest quality factors, between 4 and 5 $kPa^{-1}$.

Particle charge did not affect filtration. Filtration was nearly identical for uncharged and neutralized particles (Appendix A), consistent with Zangmeister et al. (2020), who found that the effect of particle charge is minimal for cloth-based masks, at a smaller challenge aerosol size.

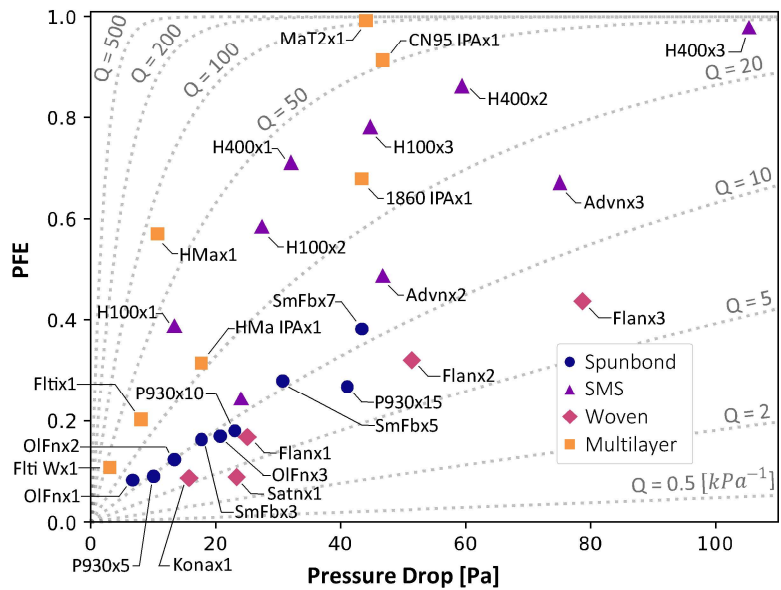

**Fig. 3.** Particle filtration efficiency at 1.0 μm versus pressure drop for all sample materials. In the legend, "SMS" denotes spunbond-meltblown-spunbond, and "Standard" denotes standard-compliant disposable. Samples noted with "IPA" were treated with isopropanol. The number of layers is denoted by the suffix, for example, "x2" being 2 layers.

## 3.2 Homogeneous layering

For a subset of the materials, the effect of layering was considered. Results for each material showed consistent quality factors across different numbers of layers in the test sample (Supplemental Information Figure C1,C2,C3 and Supplemental Information D). Some recent studies (Zangmeister et al., 2020; Zhao et al., 2020) deviated from this model, with higher measured filtration in the first layer than subsequent layers. Possibly this discrepancy results from their focus on smaller particles, which are more influenced by electrostatic effects.

Irregular variations in pressure drop, PFE and quality factors (Figure 4; Supplemental Information Figure D1) are largest in the craft-grade spunbond materials (Oly Fun, SmartFab, Pellon 930). Backlit microscopy showed variations of fibre density in localized regions (Supplemental Information Figure D2) that likely underlie the measurement variations.

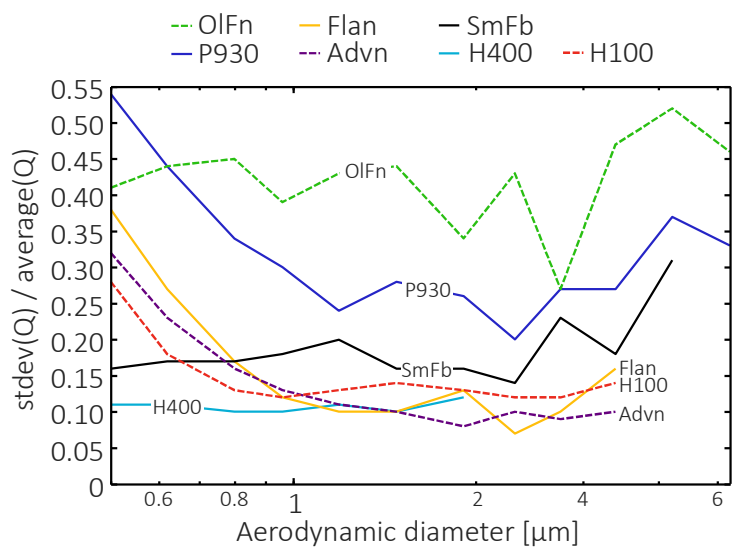

**Fig. 4.** For each material StDev(Q) / Average(Q) is from 18 samples, with 6 samples per layer count and 3 different numbers of layers.

### 3.3 Heterogeneous layering

All seven multilayer combinations had a hydrophilic skin-facing layer and all but G had a hydrophobic outer layer, as recommended by the WHO. Flannel's high-quality factor and comfort make it a desirable hydrophilic layer, however, with a high pressure drop, it reduces the acceptable pressure drop for the filtering layer. Since polypropylene and polyester are also hydrophobic, all the nonwovens (SMS and SB) are appropriate for the hydrophobic layer(s).

Table 2. Material sets evaluated as candidates for masks. Material details are in Table 1.

| Set | Layers | | |
|---|---|---|---|
| A | Advancheck SM Sx2 | | Flannel x1 |
| B | Advancheck SMS x3 | | Sateen x1 |
| C | SmartFab Thick x1 | H100 x2 | Flannel x1 |
| D | Filti x1 | | Flannel x1 |
| E | Pellon 930 x2 | Type 2 Red Cross Surgical Mask x1 | Flannel x1 |
| F | SmartFab Thick x3 | | Flannel x1 |
| G | Kona x1 | | Flannel x2 |

Figure 5 indicates that, for all combinations except D, the effect of layering is well described using Eqs. (3) and (4). This supports the idea that layers act independently for both homogeneous and heterogeneous sets. Combination D, which uses Filti, will be discussed below.

Although the measured and calculated estimates for most combinations were quite similar, the measured filtration of each individual Filti sample (PFE of 13 to 17%) was significantly worse than the combination of Filti and flannel (PFE of 92 to 97% at 0.5 μm), resulting in a significant discrepancy between the calculated (red, dotted lines) and the measured (red, solid lines) filtration for set D. Ballard et al. (2021) also observed variability in Filti's single-layer filtration ranging from 80% to 95% at 0.3 μm, but our measured single-layer Filti filtration of 13% to 17% at 0.5 μm was consistently and significantly worse. The root of the variability is unknown but was shown to be dependent on the sheet source used as samples from the same sheet showed consistent PFE.

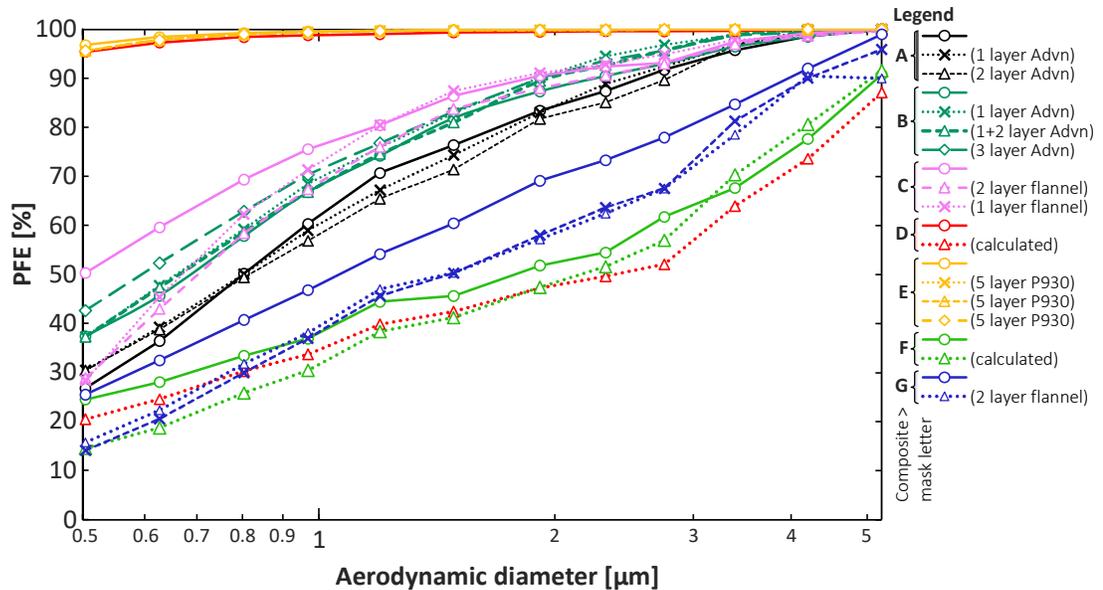

**Fig. 5.** Measured filtration overlaid with theoretical filtration from homogenous layering testing. For each color the solid line represents measured values while the dotted lines represent the theoretical values.

**3.4 Effect of machine washing**

The effect of machine washing and drying on the heterogeneous sets A-G (see Table 2) is summarized in Figure 6. For larger particles (4.2 μm), all sets showed good filtration efficiency before and after washing (PFEs of 72% to 100%). After 10 cycles of the washing protocol, sets C and D had a statistically significant decrease in filtration efficiency ($\alpha = 0.1$ in a two tailed t-test) at 4.2 μm, while all other sets showed no statistically significant difference.

For 0.5 μm particles, sets D and E had the best filtration efficiency (92%-97%, 96%-98%, respectively) before washing. All sets decreased in filtration efficiency after the washing protocol, with sets C, D and E degrading most drastically (post-wash filtration 18%-20%, 36%-42%, 33%-40%, respectively). This filtration degradation may result from the wash and IPA soak neutralizing electrostatic charge in the meltblown cores of H100 and Type 2 Red Cross Surgical Mask and in the proprietary nanofiber technology in Filti. Sets A, B, F and G showed moderate decrease in filtration efficiency at 0.5 μm.

Set F with spunbond nonwoven polypropylene was the most breathable set with a significantly lower pressure drop than all the other combinations. Sets C, D, and E showed a significant increase in pressure drop after washing, possibly due to the change in porosity and fiber density post-wash. The outer layers of sets D and E (Filti and Pellon 930, respectively) showed visual deterioration after wash. Sets A, B, and F showed small changes in pressure drop; the calendered spunbond nonwoven in these sets (Advancheck SMS and SmartFab) may have provided stability to the fibers. Set G (woven) decreased in pressure drop, perhaps due to an increased pore size after the washing protocol.

Accounting for both filtration and pressure drop, the quality factor of sets C, D, and E started out significantly higher than the other sets, but this performance advantage largely disappeared after washing.

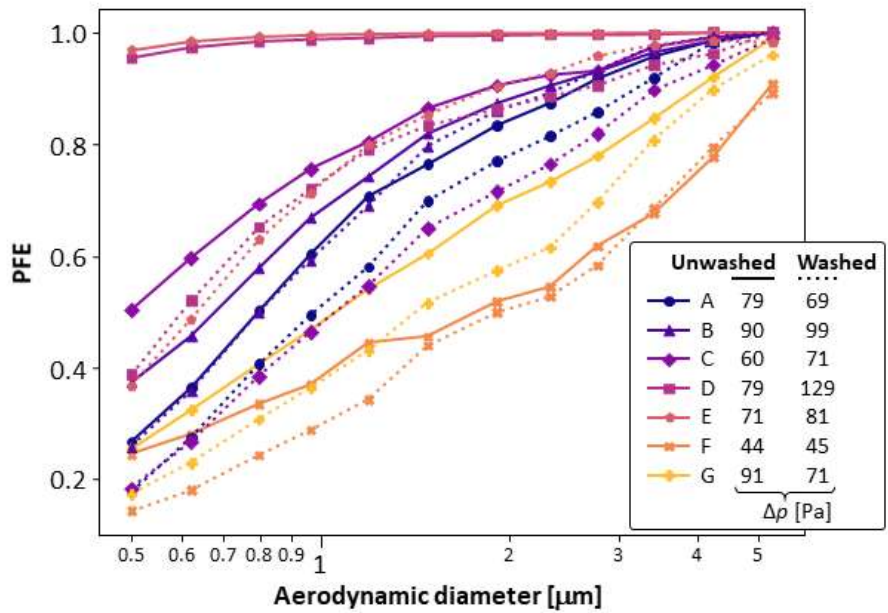

**Fig. 6.** Measured filtration for washed (dotted) and unwashed (solid) masks. Colors and symbols denote the different mask sets as described in Table 2. Pressure drop is indicated in the legend.

# 4 CONCLUSIONS

Washable fabric masks remain in widespread use in many parts of the world and offer a sustainable alternative to disposable masks. We evaluated several families of candidate materials before and after machine washing and drying, measuring the pressure drop and the filtration of particles between 0.5 and 5 μm aerodynamic diameter, and compared materials based on quality factor, a metric combining filtration with pressure drop.

Cottons (Kona, sateen) had the lowest quality factor, with cotton flannel exhibiting a better quality. However, the non-woven materials had the highest performance. Before washing, samples containing meltblown polypropylene (SMS surgical wraps, N95s, and surgical masks) outperformed spunbond polypropylene (OlyFun, SmartFab) and polyester (Pellon 930), especially for submicron particle filtration. For the nonwoven craft materials, we observed fiber density variations in localized regions, resulting in variation in the measured pressure drop and filtration efficiency between samples. The filtration for Fitli was very different when measured in individual samples versus within a heterogenous multi-layer set, with unknown reasons for the large variation.

After machine washing and drying, and an IPA soak, the samples with meltblown materials degraded to similar quality factors as the spunbond materials. Degradation was greatest for filtration of the smallest particles (0.5 μm), where electrostatic forces are of key importance in electret-based filters such as the meltblown layers of an N95. Possible degradation mechanisms include neutralizing the quasi-static electric charge of the meltblown in the first wash, with subsequent washes removing or melting fine fibers. Encasement between layers of spunbond did not protect the meltblowns enough to maintain filtration excellence through the machine wash and dry cycles used in our study. Gentler wash methods can reduce degradation (Everts et al., 2021). The higher quality factor of flannel, relative to woven cotton, may be due to its relatively disordered and fluffy structure, visually like that of the non-woven materials in this study.

With the exception of the set containing Filti, the measured penetration and pressure drop of the homogeneous and heterogenous multi-layer sets were consistent with the simple theory that the net penetration is the product of constituent layer penetrations and the net pressure drop is the sum of constituent layer pressure drops.

Overall, we have shown that spunbond non-wovens and cotton flannel offer a sustainable improvement over the widely-used, woven cotton masks for scenarios in which N95 respirators are not used.

## ACKNOWLEDGMENTS

This work was partially supported by the NSERC Discovery Grant for S. Rogak. Stephen Salter of Farallon Consultants Limited provided guidance on the study formulation and direction.

**DISCLAIMER**

# Supplemental information for "Filtration and breathability of nonwoven fabrics used in washable masks"

Thomas W. Bement[1], Ania Mitros[3], Rebecca Lau[2], Timothy A. Sipkens[1,2], Jocelyn Songer[2], Heidi Alexander[1], Devon Ostrom[4], Hamed Nikookar[1], Steven. N. Rogak[1,*]

[1] *Department of Mechanical Engineering, University of British Columbia, Vancouver BC Canada*
[2] *Metrology Research Center, National Research Council of Canada, Ottawa ON Canada*
[3] *MakerMask Group, https://makermask.org*
[4] *Artist/researcher, Ostrom.ca, Toronto, ON, Canada*

[*] Corresponding author. Tel: 1-604-822-4149; Fax: 1-604-822-2403
E-*mail address*: rogak@mech.ubc.ca

## A. EFFECT OF PARTICLE CHARGE:

The DMA column used to test the effect of particle charge was the TSI 3081A Long with a rod charge of 6 kV and no sheath flow (Figure A1). This theoretically produces purely neutral particles, as the charged particles are precipitated out in the DMA column.

The method of charge neutralization specified in the NIOSH standard results in a bipolar quasi-neutral charge distribution. We followed this aspect of the standard but wanted to quantify the effect this had on the filtration efficiency. We repeated the multilayered tests with both neutralized and uncharged particles. The tests with neutral particles were done with the DMA column in line, with the sheath flow and voltage off. The tests with uncharged particles had the DMA column in line with the sheath flow and voltage at -6000 V. A two-tailed t-test, with an alpha of 0.1, showed no statistically significant pressure difference between neutralized and uncharged particles as expected.

A two-tailed t-test, comparing the means of the quasi-neutral charged vs. uncharged particles at 4.220 µm, showed no statistically significant difference in filtration efficiencies in all cases except 5 layers SmartFab Thick, 2 layers H100 and 5 layers Pellon 930. Additionally, at the smallest measured size of 0.498 µm, none of the materials show a statistically significant difference. These results suggest that the tested materials filter by mechanical modes, rather than electrostatic forces. This means the removal of charges on the filtering media from cleaning with isopropyl alcohol (IPA), likely shouldn't reduce the filtration efficiency of the materials tested.

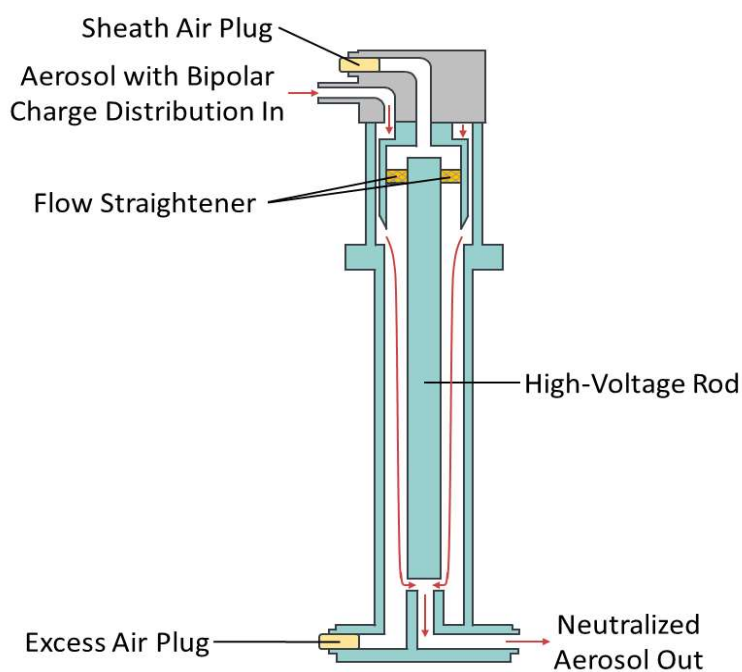

**Figure A1.** TSI 3081A Long DMA column.

Table A1. Result summary of t-Test at 4.2 µm. α=0.1, $t_{reject}$= 2.132, CI = 95%.

| Mask | T | DOF | Condition |
|---|---|---|---|
| SmFbx3 | -0.3244 | 4 | Fail to Reject |
| SmFbx5 | 2.5661 | 4 | Reject |
| SmFbx7 | 2.0468 | 4 | Fail to Reject |
| OlFnx1 | -0.1596 | 4 | Fail to Reject |
| OlFnx2 | 0.2507 | 4 | Fail to Reject |
| OlFnx3 | -0.4101 | 4 | Fail to Reject |
| Advnx1 | 1.4144 | 4 | Fail to Reject |
| Advnx2 | 0.481 | 4 | Fail to Reject |
| Advnx3 | -0.0052 | 4 | Fail to Reject |
| H400x1 | 1.2247 | 4 | Fail to Reject |
| H400x2 | -0.1459 | 4 | Fail to Reject |
| H400x3 | 1.2247 | 4 | Fail to Reject |
| H100x1 | -0.6072 | 4 | Fail to Reject |
| H100x2 | -3.678 | 4 | Reject |
| H100x3 | -1.2247 | 4 | Fail to Reject |
| P930x5 | 4.7844 | 4 | Reject |
| P930x10 | 1.8438 | 4 | Fail to Reject |
| P930x15 | 0.4232 | 4 | Fail to Reject |
| Flanx1 | 1.813 | 4 | Fail to Reject |
| Flanx2 | -0.3536 | 4 | Fail to Reject |
| Flanx3 | -0.1931 | 4 | Fail to Reject |

## B. OPS SIZE BIN AND AERODYNAMIC DIAMETER RELATION

The TSI3330 allows the user to introduce a refractive index, resulting in a corrected size for the size bin limits. In theory, this provides the physical diameter (as would be measured by microscopy). Taking the geometric mean (square root of the product) of the upper and lower bounds corresponding to a channel, we obtain the 4$^{th}$ column of Table B1. Finally, using a bulk density for solid sodium chloride of 1900 kg/m$^3$, the aerodynamic diameter associated with a particular OPS channel count is obtained. Thus we report the filtration efficiency for the smallest channel at 0.498 microns. Given uncertainties on the precise shape, refractive index and density of the sodium chloride particles, the uncertainty in the reporting size is substantial – perhaps 15%. This would translate the particle filtration data to the left or right but would have no influence on the comparisons between materials.

Table B1 OPS size bin conversion to aerodynamic diameter at all size bins

| OPS Channel | Nominal Bin Lower Limit | Refractive Index Corrected Lower Limit | Geometric Mean Diameter | Aerodynamic Diameter |
|---|---|---|---|---|
| 1 | 0.300 | 0.306 | 0.340 | 0.498 |
| 2 | 0.370 | 0.381 | 0.430 | 0.620 |
| 3 | 0.460 | 0.484 | 0.540 | 0.796 |
| 4 | 0.570 | 0.596 | 0.680 | 0.962 |
| 5 | 0.710 | 0.769 | 0.840 | 1.190 |
| 6 | 0.880 | 0.932 | 1.050 | 1.480 |
| 7 | 1.090 | 1.200 | 1.360 | 1.910 |
| 8 | 1.350 | 1.550 | 1.660 | 2.320 |
| 9 | 1.680 | 1.780 | 1.980 | 2.760 |
| 10 | 2.080 | 2.200 | 2.440 | 3.400 |
| 11 | 2.580 | 2.720 | 3.040 | 4.220 |
| 12 | 3.200 | 3.400 | 3.780 | 5.250 |
| 13 | 3.960 | 4.210 | 4.690 | 6.490 |
| 14 | 4.920 | 5.220 | 5.860 | 8.120 |
| 15 | 6.100 | 6.590 | 7.230 | 10.000 |
| 16 | 7.560 | 7.950 | 8.840 | 12.200 |

## C. FILTRATION EFFICIENCY AND PRESSURE DROP FOR INDIVIDUAL MATERIALS

Filtration efficiency was measured for all 16 OPS bins, but for compactness we present only results for 0.5 microns, 1 micron and 4.2 microns. Error bars on the following plots represent the range for (usually) 3 trials. Layers are filtering independently when materials follow lines of constant $Q$ as the number of layers are changed (eg. P930x15 has 15 layers). For all materials, filtration efficiency increases with particle sizes. Thus, most of the non-woven fabrics have $Q$ ranging from 4-12 kPa$^{-1}$ at 0.5 microns, 6-15 at 1 micron, and 30-60 at 4.2 microns.

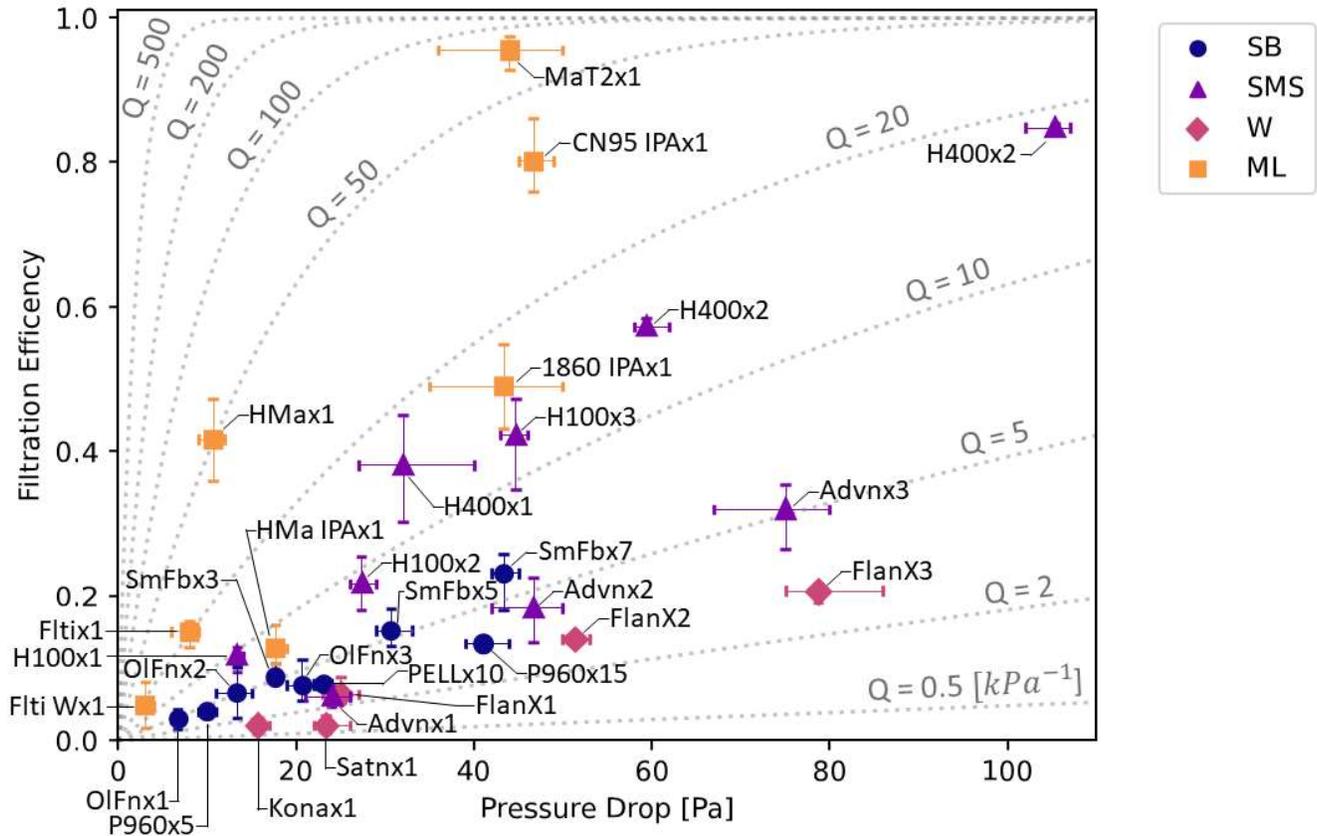

**Figure C1**. Filtration at 0.5 micron aerodynamic diameter and pressure drop.

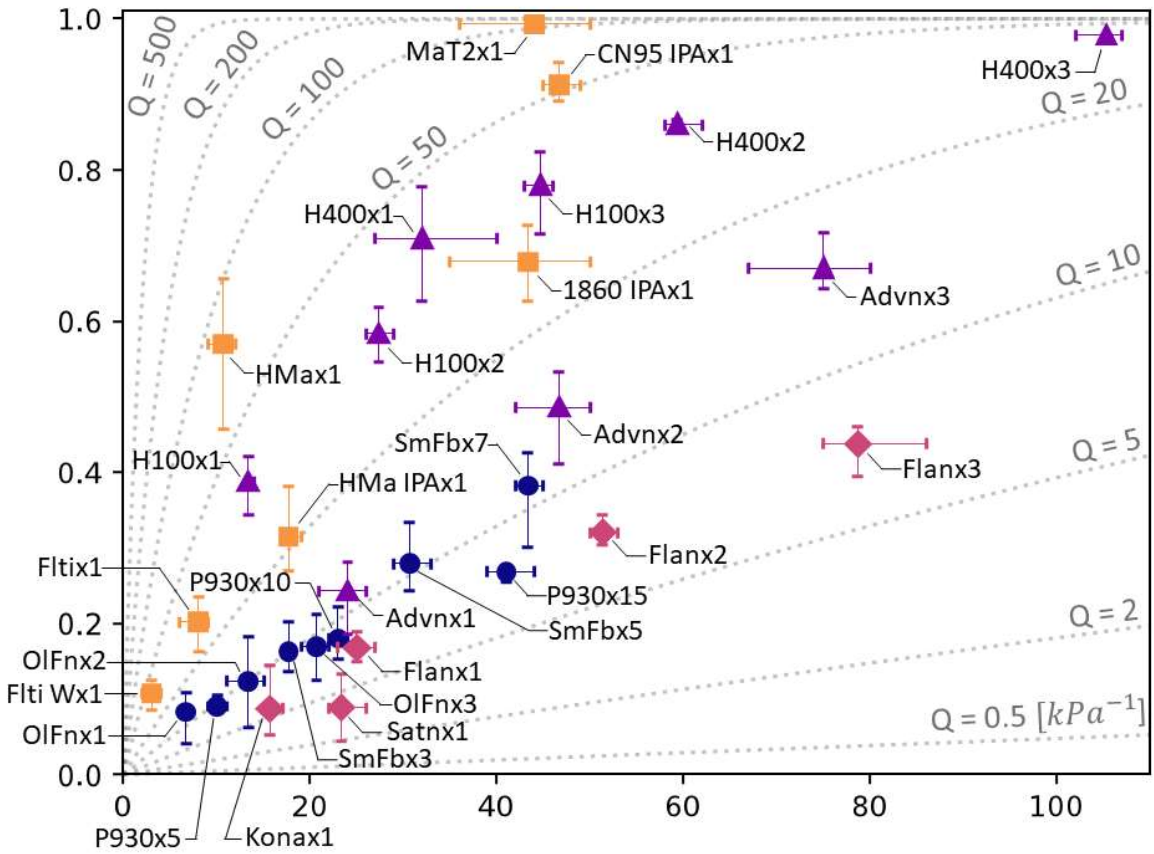

**Figure C2.** Filtration at 1.0 micron aerodynamic diameter and pressure drop. Legend for material types is given in Figure C1.

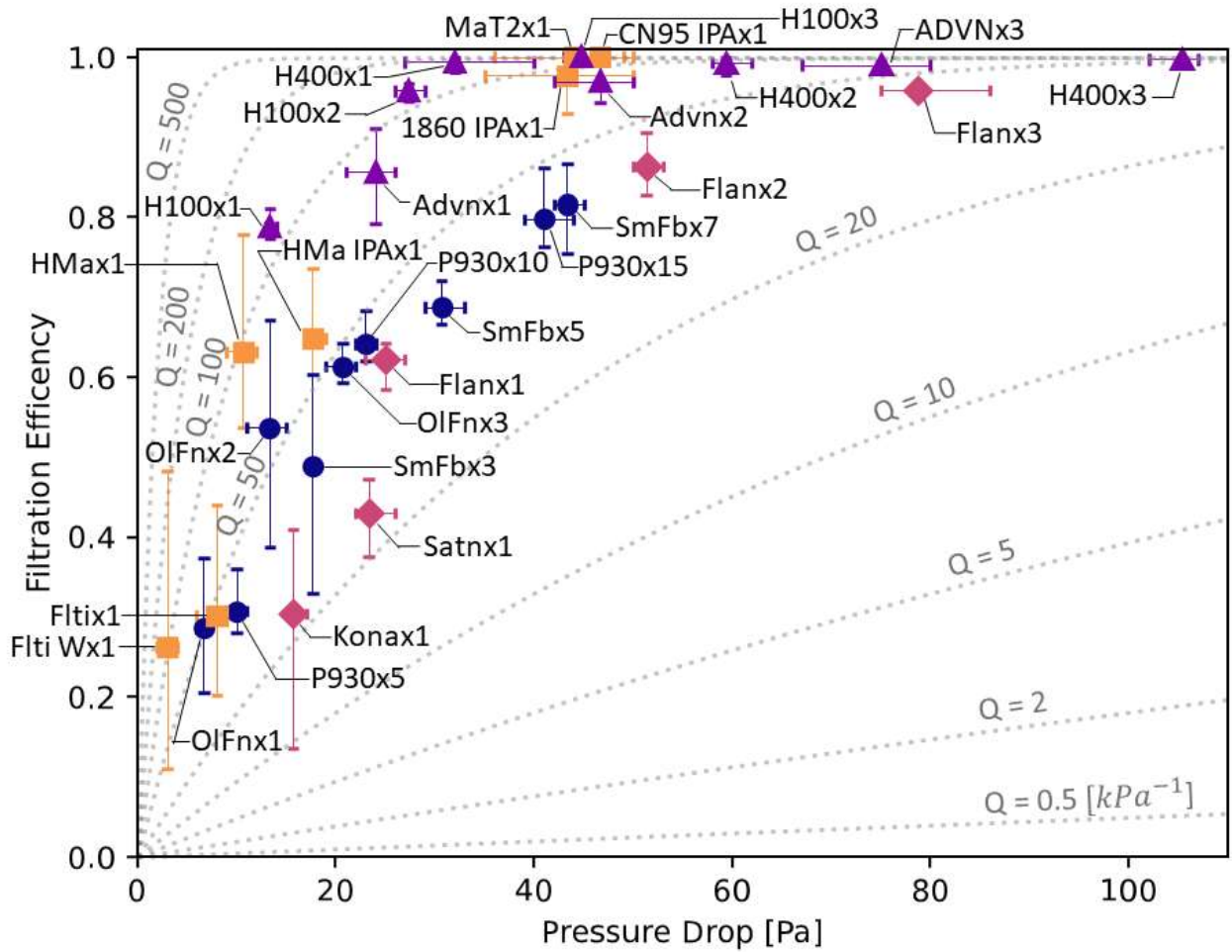

**Figure C3.** Filtration at 4.2 micron aerodynamic diameter and pressure drop. Legend for material types is given in Figure C1.

# D. MULTILAYER MATERIAL TESTING OF HOMOGENEOUS MATERIALS AND MASK ASSEMBLIES

SMS materials having high single layer filtration efficiencies have a much steeper slope because of the $P_0$ term (equation 4). Flannel also performs comparatively well with SMS materials. The Spunbond materials have slower sloped curves because of their lower single-layer filtration efficiencies.

The fitted efficiency curve (Figures D1b-D1c) doesn't exactly line up with the measured single layer efficiency. Electrostatic effects are unlikely to be responsible for the discrepancy, since 5-layer tests of Pellon 930 with the DMA column indicated negligible charge effects. Sample inconsistency likely underlies both the discrepancy between model versus measurement as well as the large error bars for OlyFun and SmartFab.

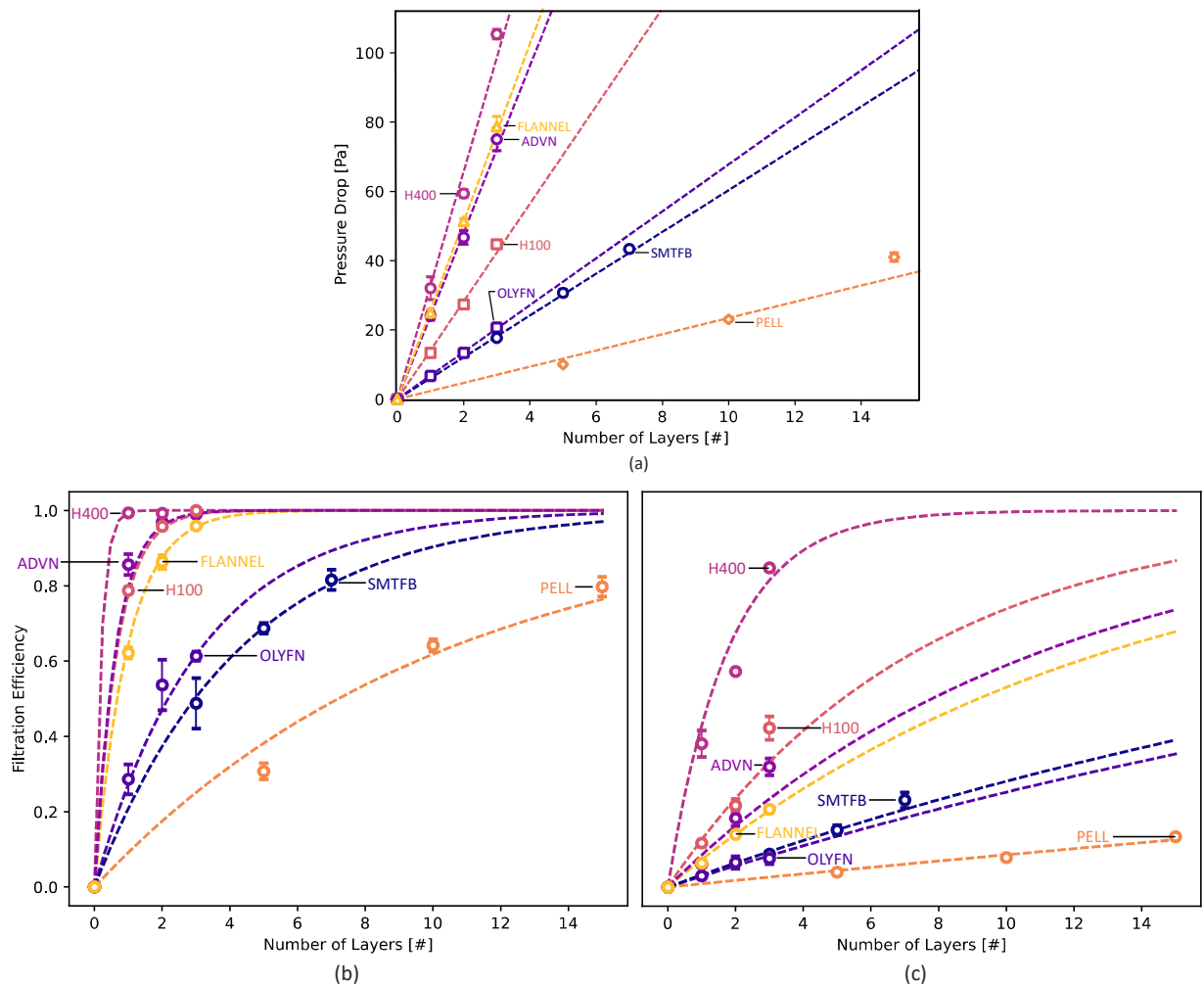

Figure D1. Pressure drop (a), filtration efficiency at 0.498 µm (b) and 4.220 µm (c) based on number of layers for homogeneous multilayer material stacks. Circles represent average filtration measured from 3 samples. Curves are calculated by applying a non-linear least squares to a specified fit function (equation 4). Error bars represent one standard error based off the three samples.

Visual inspection shows qualitatively different regions in these materials (Figure D2).

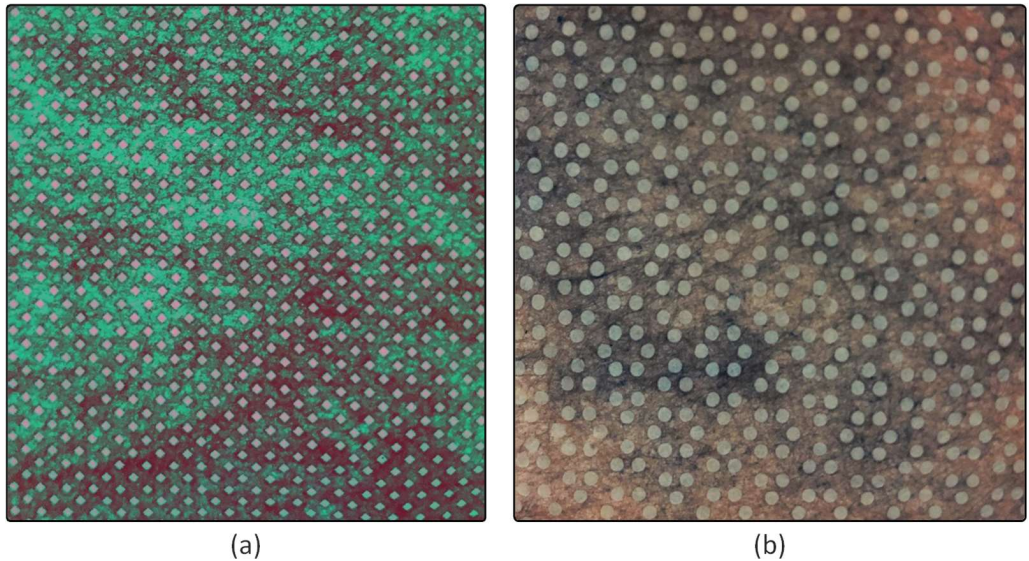

(a)            (b)

Figure D2. Photo of OlyFun (a) and SmartFab (b) materials with enhanced color balance. Visual variation in light levels passing through the sample is observable. Light green represents less dense areas while darker regions are more densely packed.

Applying the models acquired when fitting the homogeneous material tests and applying the equations used for predicting the pressure drop and penetration assuming each layer behaves independently a few potential mask stacks can be characterized based on their theoretical combined performance and their measured performance (Figures D3a-D3b).

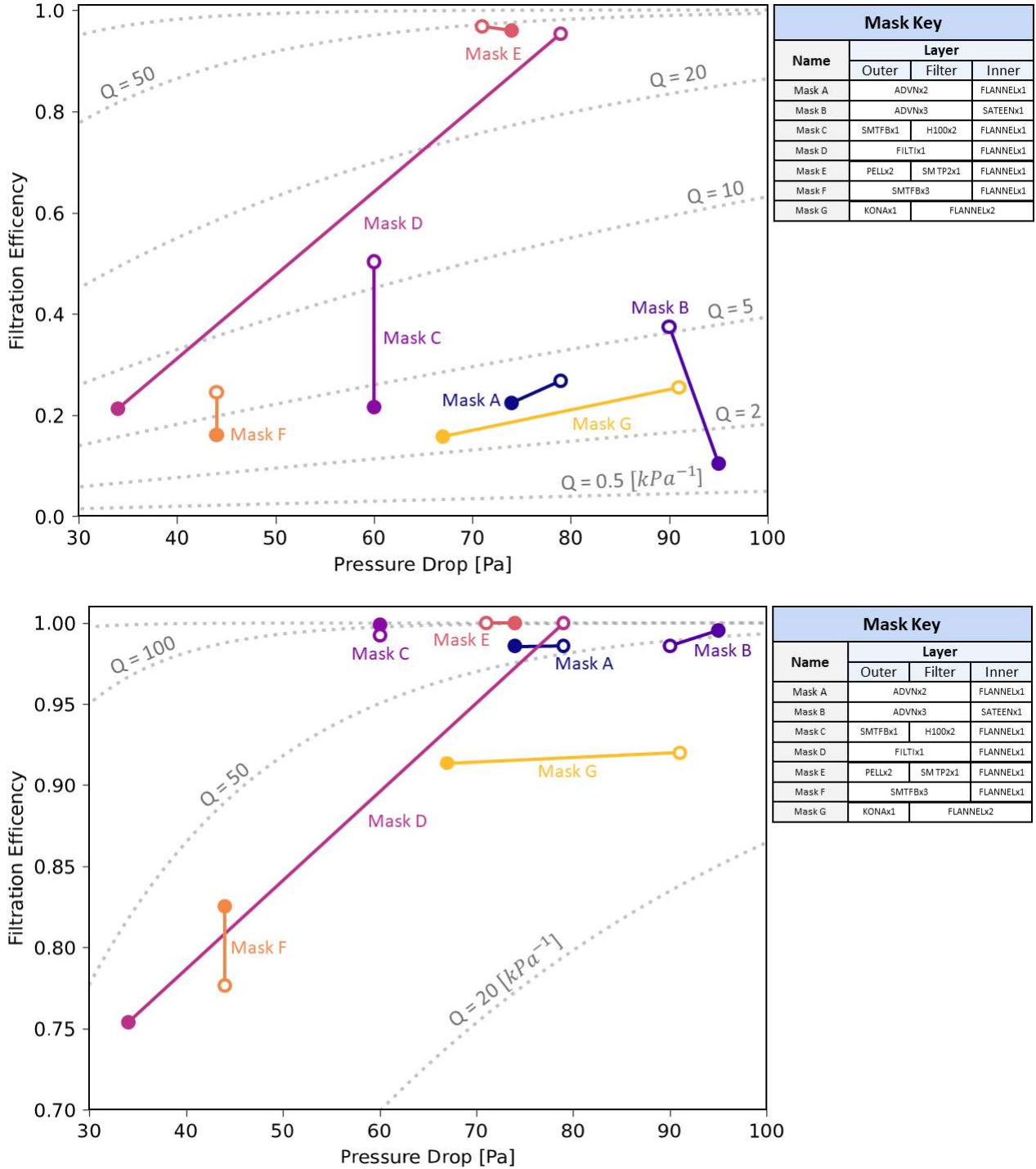

**Figure D3.** Model predictions (closed symbols) at 0.5 µm (top) and 4.2 µm (bottom) versus experimental values (open symbols) for proposed mask stacks of Table 2. The large difference for Mask D is due to the single layer efficiencies of FILTI. These values were much lower when tested individually rather than with other materials.